\documentclass[prd,superscriptaddress,showpacs,floats,nofootinbib,preprint,floatfix]{revtex4}

\usepackage{amsmath,amssymb}
\usepackage{graphicx}
\usepackage{bm}

\begin{document}

\title{Quarkonia in the deconfined phase: effective potentials and lattice
  correlators} 
\author{W.M. Alberico}
\affiliation{Dipartimento di Fisica Teorica dell'Universit\`a di Torino and \\ 
  Istituto Nazionale di Fisica Nucleare, Sezione di Torino, \\ 
  via P.Giuria 1, I-10125 Torino, Italy}
\author{A. Beraudo}
\affiliation{Dipartimento di Fisica Teorica dell'Universit\`a di Torino and \\ 
  Istituto Nazionale di Fisica Nucleare, Sezione di Torino, \\ 
  via P.Giuria 1, I-10125 Torino, Italy}
\affiliation{SPhT, CEA-Saclay, 91191 Gif-sur-Yvette, France}
\author{A. De Pace}
\affiliation{Dipartimento di Fisica Teorica dell'Universit\`a di Torino and \\ 
  Istituto Nazionale di Fisica Nucleare, Sezione di Torino, \\ 
  via P.Giuria 1, I-10125 Torino, Italy}
\author{A. Molinari} 
\affiliation{Dipartimento di Fisica Teorica dell'Universit\`a di Torino and \\ 
  Istituto Nazionale di Fisica Nucleare, Sezione di Torino, \\ 
  via P.Giuria 1, I-10125 Torino, Italy}

\begin{abstract}
The Schroedinger equation for the charmonium and bottomonium states at finite
temperature is solved by employing an effective temperature dependent potential given by a linear combination of the color singlet free and internal energies 
obtained on the lattice from the Polyakov loop correlation functions. 
The melting temperatures and other properties of the quarkonium states are 
evaluated. The consistency of the potential model approach with the available 
lattice data on the quarkonium temporal correlators and spectral functions is
explored. 
\end{abstract}


%\date{\today}

\maketitle

\section{Introduction}

The anomalous suppression of the $J/\psi$ production in heavy ion collisions
--- experimentally observed \cite{Abr00,Ale05} in the depletion of the dilepton
multiplicity in the region of invariant mass corresponding to the $J/\psi$
meson ---has been proposed long time ago as a possibly unambiguous signal 
of the onset of deconfinement \cite{Mat86}. Indeed, in
Ref.~\cite{Mat86} it was argued that charmonium states can only be
produced in the first instants after the nucleus-nucleus collision, before the
formation of a thermalized Quark-Gluon Plasma (QGP). 
Then, in their path through the deconfined medium, the original $c\bar{c}$
bound states tend to melt, since the binding (colour) Coulomb potential is
screened by the large number of colour charges.  
This, in turn, would produce an anomalous (with respect to normal nuclear
absorption) drop in the $J/\psi$ yields. 
In this picture it was implicitly assumed that, once the charmonium
dissociates, the heavy quarks hadronize by combining with light quarks only
(recombination leading to a secondary $J/\psi$ production being neglected). 

Since in hadronic collisions a sizable fraction of the measured $J/\psi$'s 
comes from the decay of excited charmonium states (bottomonium feed-down 
contribution being negligible at the Super Proton Synchrotron (SPS) conditions)
and the latter are expected to dissociate at lower temperatures, a mechanism of
{\em sequential suppression} has been  proposed, with the aim of reproducing
the $J/\psi$ suppression pattern as a function of the energy density reached in
the heavy ion collision (the highest temperatures and energy densities being
reached in the most central collisions) \cite{Gup92,Dig01a,Dig01b,Kar05,Kha97}.

In all the above mentioned works the reason for charmonium melting in a hot
environment was essentially found in the Debye screening of the potential among
the two color charges. In this regard one immediately faces a few theoretical 
problems.

First of all, it is not even clear, at present, whether a temperature-dependent
potential can describe medium modifications of quarkonia, although there are
efforts aiming to derive a potential at finite temperatures from the underlying
QCD (see, e.~g., Refs.~\cite{Sim05,Lai06}). 
Pragmatically, one can follow a phenomenological approach, building an
effective model and testing it against quarkonium properties directly
calculated in QCD. However, even assuming the validity of the potential model
at finite temperatures, one has to tackle the problem of finding the
appropriate effective screened potential to insert into the Schroedinger
equation, as it will be discussed below.

Indeed, when the mechanism of anomalous $J/\psi$ suppression was originally
proposed, lattice calculations could not provide any quantitative estimate of
the Debye screening length and the authors had to employ a schematic model for
the $c\bar{c}$ interaction. 
Nowadays very precise lattice calculations of Polyakov line correlators are
available, for different numbers of light dynamical fermions
\cite{Kac02,Kac04a,Kac04b,Kac05,Pet04}. 
It has been known for a long time \cite{McL81} that from these correlators it
is possible to extract --- in hot QCD, as a function of the temperature and of
the $Q\bar{Q}$ separation --- the change in free energy once a $Q\bar{Q}$ pair
is placed in a thermal bath of gluons and light quarks.

The free energy obtained in these calculations, taken in the color singlet
channel, has been used as a temperature-dependent potential and inserted into
the Schroedinger equation in a number of works
\cite{Dig01a,Dig01b,Won02a,Won02b}. 
Such a choice turned out to give very low melting temperatures for all the
charmonium states: the dissociation temperatures $T_d=1.10T_c$ \cite{Dig01b}
and $T_d=0.99T_c$ \cite{Won02a} were found for the $J/\psi$, all the other
charmonium states melting well below $T_c$.

On the other hand, since the free energy contains an entropy contribution, it
was soon realized that employing the change in {\em internal} (instead of 
{\em free}) energy as an effective $Q\bar{Q}$ potential could appear better
justified \cite{Shu04,Alb05,Sat06a,Sat06b}. This choice results in a more
attractive potential, leading to a melting temperature for the $J/\psi$ around 
$1.5\div2T_c$, the other charmonium states ($\psi'$ and $\chi_c$) dissociating
a bit above $T_c$. 
These findings appear in agreement with (quenched) lattice spectral function
studies \cite{Dat04,Asa03,Asa04,Ume05,Jak06}, at least for what concerns the 
melting temperatures of charmonia. Lattice results are starting to be available
also for the bottomonium spectral function \cite{Pet05,Dat06,Jak06} and for the
unquenched case \cite{Aar06}. They appear also able to explain the most recent 
analysis of the Na50 (Pb-Pb), Na60 (In-In) \cite{Lou06} and RHIC (Au-Au) 
\cite{Sat06b} data on $J/\psi$ production, which seems to favour a scenario in
which only the excited states of charmonium are anomalously suppressed.

Other choices for the effective potential can be found in the literature
\cite{Won05,Won06a,DiG05}. In Refs.~\cite{Won05,Won06a}, in particular, the 
author argues that the internal energy obtained from lattice data through the
standard thermodynamical relation contains also a gluon and light quark
contribution. A subtraction procedure is constructed, leading to an effective
$Q\bar{Q}$ potential given by a linear combination of the lattice free and
internal energies (normalized to the case with no heavy color sources), with
temperature dependent coefficients obtained from the QCD equation of state.

Results for the quarkonium dissociation temperatures, in
potential models based upon lattice data, appear to be consistent with
independent lattice results from spectral functions calculations.
Indeed, because of the huge numerical effort required by lattice spectral
studies, a very fine temperature scan is not available yet.
Furthermore, since --- as mentioned above --- a derivation of the potential
model at finite temperature from first principles is still lacking, it would be
desirable to test it against other observables.

Actually, full information on the fate of the $Q\bar{Q}$ states in thermal
equilibrium in a hot environment is encoded, depending upon the channel one is
considering, in the meson spectral functions. 
The latter can be extracted from the euclidean
temporal correlators of mesonic currents (measured on the lattice) by inverting
an integral transform. Such a task is usually achieved with a technique
referred to as Maximum Entropy Method (MEM) \cite{Asa01,Nak99} and lattice
results using this method can be found in
Refs.~\cite{Dat04,Asa03,Asa04,Ume05,Jak06,Pet05,Dat06,Aar06}.

Since the reliability of MEM to extract the spectral functions is not fully
established yet, the check of consistency of the potential model has been
mainly devoted to a direct comparison with the euclidean correlators in a
number of papers \cite{Moc05,Moc06a,Moc06b,Moc06c,Won06b,Cab06}. 
In these works the authors, starting from different screened potentials,
calculate in a given channel the corresponding charmonium spectral function, 
which gets contributions both from bound states (as long as they are
supported by the potential) and (starting from a threshold energy) from
scattering states. Convoluting the spectral function with a thermal kernel,
they eventually obtain the charmonium correlator along the imaginary temporal
direction. Such a quantity can then be compared with the ones measured on the
lattice. In Refs.~\cite{Moc05,Moc06a,Moc06b,Moc06c} the authors
point out that this procedure leads to a disagreement between lattice and
potential model results. Hence they conclude that a study of the $Q\bar{Q}$
states in the QGP in terms of screened potentials may not be able to catch the
right physics. On the other hand, the authors of Ref.~\cite{Won06b} propose of
curing the discrepancy (at least in the pseudoscalar channel) by keeping in the
continuum part of the potential model spectrum only the resonant contributions,
i.~e. by subtracting the free gas states. We believe that this procedure might
be incorrect, since the evaluated correlator has to be compared with the
lattice ones, which do have a free gas (infinite temperature) limit.

The comparison between lattice data and potential model calculations is usually
done by considering a quantity constructed ad hoc, to display the temperature
dependence of quarkonium properties, namely the ratio between euclidean
correlators above and below the critical temperature (see next Section). 
In a potential model this ratio turns out to be very sensitive to the treatment
of the continuum, e.~g. to the threshold energy.
This is clearly a problem for the calculation of correlators below the critical
temperature, since, for instance, at $T=0$ the potential model is based on
confining potentials and the continuum spectrum has to be added by hand.
Secondly, lattice calculations appear to be dominated by finite size
artifacts right in the continuum region, leading to unphysical peaks in
the spectral functions, even in the infinite temperature
limit~\cite{Kar03,Aar06}. 
Moreover, the asymptotic high energy behavior, $\sigma(\omega)\sim\omega^2$,
of the continuum spectral functions is not reproduced by lattice calculations,
since the finite lattice spacing provides an ultraviolet cutoff.

In this paper, following Refs.~\cite{Alb05,Won05,Won06a}, we extract from the
lattice data for Polyakov line correlators an effective temperature-dependent
$Q\bar{Q}$ potential and we use it in order to understand to which extent a
comparison with mesonic temporal correlators obtained on the lattice can be
pursued. 
In particular, we study  the effect of different models for the continuum and
we try to keep under control the uncertainties introduced by the need of
calculating correlators below $T_c$ by employing the effective potential
derived from lattice data for $T<T_c$ as well.

We wish to stress how important (both from a theoretical point of view and for
its experimental consequences) would it be to show that an approach in terms of
screened potentials is indeed able to catch the main features of the behavior
of quarkonia in a hot environment. 

Indeed, at present, from finite temperature lattice meson spectral functions
one can extract important information on the fate of the ground state in the
different quantum number channels, concerning its position, strength and
melting temperature. Unfortunately, no reliable information can be extracted so
far on the excited states, due to the difficulty of disentangling their
peaks (if any) from lattice artifacts. Although this is not a severe
limitation for charmonium studies, where one expects just a few states to
(possibly) survive after the phase transition ($J/\psi$, $\chi_c$ and $\psi'$,
the latter being the only excited state), this is no longer true for the
bottomonium.

From the experimental point of view this turns out to be a strong limitation,
especially in view of the ALICE experiment, where one expects a sizable
production of $b\bar{b}$ pairs. In particular the vector channel states will be
experimentally accessible through their electromagnetic decays into dilepton
pairs \cite{ALI06}.
Moreover, charmonium yields will also get a feed-down contribution from 
$b\bar{b}$ states.
The knowledge of the melting temperature of the different states is then
important in order to develop a reliable model of sequential suppression to be
eventually compared with the experimental data. 

In this regard, it is clearly of importance, in view of quantitative studies of
bottomonia above $T_c$, to demonstrate that potential model calculations are
consistent with the results obtained from the lattice spectral functions.

Although, as we shall see, definite conclusions cannot be drawn yet, our work
constitutes a promising consistency check of the potential model, given the
present status of lattice calculations at finite temperatures.
Of course, the final aim should be, on the one hand, to establish a clean
theoretical justification of the potential model and, on the other hand, to
employ it in the interpretation of the next generation of heavy ion
experiments. 
As a general remark, we remind the reader that potential model approaches, like
the one adopted in this paper, give a {\em static} picture of the behavior of
charmonium in the QGP, the effects of the interaction with the deconfined
quarks and gluons being encoded in the {\em screened} $Q\overline{Q}$
interaction. 
Alternative pictures are available in the literature, which attempt to predict
final charmonium yields in nucleus-nucleus collisions as arising from the
interplay of dissociation--recombination processes~\cite{Gra01,Gra02,Gra04} 
or in the framework of statistical models of 
hadronization~\cite{Gaz99,Mun00,Mun06}. 

The paper is organized as follows: in Section~\ref{sec:formalism} we first
briefly review the formalism of finite temperature quarkonium spectral
functions and imaginary time correlators and then the procedure followed to
extract the effective $Q\bar{Q}$ potential from lattice data; in
Section~\ref{sec:results} we compare the outcome of the potential model to
results from lattice calculations, both for the spectral functions and the 
euclidean correlators; finally, in Section~\ref{sec:concl} we summarize and
discuss our results.
 
\section{Formalism}
\label{sec:formalism}

\subsection{Lattice correlators and spectral functions}
\label{subsec:lcsp}

The standard object introduced in lattice studies of quarkonium properties at
finite temperature $T$ is the Euclidean time correlator, defined as the thermal
expectation value of a hadronic current-current correlation function in
Euclidean time $\tau$ for a given mesonic channel $H$ (see, e.~g.,
Ref.~\cite{Bra05}, Chap.~7): 
\begin{equation}
\label{eq:GH}
  G_H(\tau,T) = \langle j_H(\tau) j_H^\dagger(0) \rangle,
\end{equation}
where $j_H=\bar{q}\Gamma_H q$ and $\Gamma_H=1$, $\gamma_5$, $\gamma_\mu$,
$\gamma_\mu\gamma_5$. The four vertex operators $\Gamma_H$ correspond,
respectively, to the scalar, pseudoscalar, vector and axial-vector mesonic
channels, which in turn, at zero temperature, correspond to the $\chi_{c0}$
($\chi_{b0}$), $\eta_c$ ($\eta_b$), $J/\Psi$ ($\Upsilon$) and $\chi_{c1}$
($\chi_{b1}$) quarkonium states for the $c\bar{c}$ ($b\bar{b}$) system,
respectively. Although the correlation functions can in general be defined for
any spatial momentum $p$, most lattice studies are restricted to the case $p=0$
and we shall consider, in the following, only this case.

The correlators of Eq.~(\ref{eq:GH}) are directly evaluated in lattice QCD,
whereas physical observables are related to the spectral function $\sigma_H$, 
i.~e. to the imaginary part of the real time retarded correlators. 
The two quantities are connected by an integral transform 
\begin{equation}
\label{eq:GHs}
  G_H(\tau,T) = \int_0^\infty d\omega\,\sigma_H(\omega,T) K(\tau,\omega,T),
\end{equation}
which is regulated by the temperature kernel
\begin{equation}
  K(\tau,\omega,T) = \frac{\cosh[\omega(\tau-1/2T)]}{\sinh[\omega/2T]},
\end{equation}
the energy integration extending over the whole spectrum.

Extracting the continuous spectral function from the discrete and finite --- 
and usually rather limited --- set of lattice data is an ill-posed problem.
It has been tackled through the application of the MEM \cite{Asa01,Nak99}, 
which appears to yield promising results, although its reliability has yet to
be confirmed. 

The MEM analysis has been applied by different groups to the $c\bar{c}$ system,
both below and above the critical temperature and mainly in the quenched
approximation (see, however, Ref.~\cite{Aar06} for a study in presence 
of dynamical fermions). 
The spectral functions obtained in these studies present, schematically,
similar features: a well defined peak in correspondence to the ground state
meson in a channel of given angular momentum ($S$- or $P$-wave) up to the
dissociation temperature; inability to resolve radial excitations; presence 
in the highest part of the spectrum of peaks that are associated to lattice
artifacts. 
Actually, the most meaningful information extracted from the lattice spectral
functions concerns the existence and mass value of the lowest quarkonium states
at a given temperature. 

However, the survival, e.~g., of the $J/\Psi$ up to temperatures
around $1.5\div 2T_c$ has important consequences for the interpretation of
heavy ion experiments and it has raised the need for a confirmation not
suffering from the uncertainties of the MEM analysis.
A procedure has been devised that makes directly use of the euclidean
correlators, by comparing the behavior of the correlators above and below
$T_c$ \cite{Dat04}. Specifically, one introduces the ratio between the
correlation function  $G_H(\tau,T_>)$ of Eq.~(\ref{eq:GHs}) at a temperature
$T_>>T_c$ and the so-called {\em reconstructed correlator}
\begin{equation}
  G_H^{\text{rec}}(\tau,T_>,T_<) = \int d\omega\,\sigma_H(\omega,T_<) 
    K(\tau,\omega,T_>),
\end{equation}
calculated using the kernel at $T_>>T_c$ and the MEM spectral function at some
reference temperature $T_<<T_c$. This procedure should eliminate the trivial
temperature dependence due to the kernel and differences from one in the ratio
should then be ascribed to the temperature dependence of the spectral function.
Furthermore, the MEM result for $\sigma_H(\omega,T_<)$ --- because of the 
larger number of lattice sites at low temperature --- is more robust. 
On the other hand, at high temperature the correlator is directly measured on 
the lattice, thus avoiding the extraction of the spectral function with the MEM procedure

Indeed, lattice studies show that the ratio $G_H/G_H^{\text{rec}}$ stays
around one up to the same dissociation temperatures extracted from the MEM
analysis and then it departs from one following different patterns in the
different channels \cite{Dat04,Jak06}. 

The other approach followed in the literature in the determination of the
quarkonia dissociation temperatures --- which is based on the use of effective
potentials for the quarkonium systems --- has reached similar conclusion about
the dissociation temperatures, at least when making use of effective potentials
extracted from lattice data on $Q\bar{Q}$ free energies at finite temperature
\cite{Won05,Alb05,Won06a}.

In a potential model approach one solves the Schroedinger equation for a given
potential, thus getting not only the spectrum but also the wave functions of
the $Q\bar{Q}$ pair. It is then fairly straightforward to get the corresponding
spectral function as the imaginary part of the $Q\bar{Q}$ propagator, namely
\begin{eqnarray}
\label{eq:sigmaHPM}
  \sigma_H(\omega,T) &=& \frac{1}{\pi} \text{Im} G_H(\omega) \nonumber \\
  &=& \sum_n\left|\langle 0|j_H|n\rangle\right|^2\delta(\omega-E_n) 
    \nonumber \\
  &=& \sum_n F_{H,n}^2 \delta(\omega-M_n) + \theta(\omega-s_0) 
    F_{H,\omega-s_0}^2, 
\end{eqnarray}
where $n$ represents all the relevant quantum numbers (in the second line the
sum over $n$ actually stands for both the sum over discrete states and the
integration over the continuum) and $E_n=s_0+\epsilon_n$, $s_0$ being the
continuum threshold and $\epsilon_n$ the eigenvalue associated to the state
$|n\rangle$. In the last line we have separated the discrete and continuum
contributions, introducing the couplings $F_{H,n}\equiv\langle0|j_H|n\rangle$ 
and $F_{H,\epsilon}\equiv\langle0|j_H|\epsilon\rangle$ and the bound state 
masses $M_n$ (a normalization of the discrete and continuum states as 
$\langle n|n'\rangle=\delta_{nn'}$ and 
$\langle\epsilon|\epsilon'\rangle=\delta(\epsilon-\epsilon')$, respectively, is understood).

Following Ref.~\cite{Bod95} (see also Ref.~\cite{Moc06b}) one can express the 
couplings in terms of the wave function at the origin for the $S$-states
(pseudoscalar and vector channels),
\begin{equation}
  F^2_{PS} = \frac{N_c}{2\pi} |R(0)|^2 \quad \textrm{and} \quad
    F^2_{V} = \frac{3N_c}{2\pi} |R(0)|^2,
\end{equation}
and in terms of the first derivative of the wave function at the origin for the
$P$-states (scalar and axial-vector channels),
\begin{equation}
  F^2_{S} = \frac{9N_c}{2\pi m^2} |R'(0)|^2 \quad \textrm{and} \quad
    F^2_{A} = \frac{9N_c}{\pi m^2} |R'(0)|^2,
\end{equation}
$N_c$ being the number of colors and $m$ the quark mass.

By inserting the spectral function of Eq.~(\ref{eq:sigmaHPM}) into
Eq.~(\ref{eq:GHs}), one gets the potential model expression for the Euclidean
correlators:
\begin{equation}
\label{eq:GHSc}
  G_H(\tau,T) = \sum_n F^2_{H,n} K(\tau,M_n,T) + \int_0^\infty d\epsilon \,
    F^2_{H,\epsilon} K(\tau,\epsilon+s_0,T).
\end{equation}
Some care is necessary in dealing with the continuum part of the spectral
function. Indeed, in Eq.~(\ref{eq:GHSc}) the integration over the excitation
energy can reach pretty high values, especially for small $\tau$'s, probing
regions where the non-relativistic dispersion relation is no longer valid.
Actually, the asymptotic relativistic spectral function $\sigma_H(\omega)$ that
one gets in perturbative QCD is known to be proportional to $\omega^2$ in every
channel, whereas it is easy to check that the non-relativistic solution of the
Schroedinger equation in the free case goes like, e.~g., $\omega^{1/2}$ for the
$S$-states.
The difference is due to the different phase space and it can be accounted for
by simply renormalizing the wave function with a phase space factor
\cite{Dur82}. 
In the calculations discussed below we will be using such wave functions,
renormalized to account for relativistic kinematical effects.

Note that in potential model calculations of the spectral functions, a mixed
representation has been sometimes employed, with a continuum part approximated
using the perturbative QCD expression with a threshold extracted from the
effective potential \cite{Moc05,Moc06b}:
\begin{equation}
\label{eq:sigmaHPMA}
  \sigma_H(\omega,T) = \sum_n F_{H,n}^2 \delta(\omega-M_n) + 
    \frac{3}{8\pi^2} \omega^2 \theta(\omega-s_0) f_H(\omega,s_0),
\end{equation}
where the function $f_H$ is given by leading order perturbative calculations
\cite{Kar01,Alb05b,Aar05} as
\begin{equation}
  f_H(\omega,s_0) = \left[a_H + b_H \frac{s_0^2}{\omega^2} \right] 
    \sqrt{1 - \frac{s_0^2}{\omega^2}}.
\end{equation}
At leading order, the coefficients $(a_H,b_H)$ are 
$(1,-1)$, $(1,0)$, $(2,1)$ and $(2,-3)$, for the scalar, pseudoscalar, vector
and axial-vector channels, respectively.
This form for the spectral function is affected by an obvious inconsistency 
between the bound and the continuum parts of the spectrum. Nevertheless, 
in the following we make use also of this expression in order to test the model
dependence of our results.

\subsection{Effective potentials from lattice data}
\label{subsec:effpot}

In Ref.~\cite{Alb05} we have provided a unified parametrization of the
temperature and separation dependence of the lattice data for the color-singlet
$Q\bar{Q}$ free energy $F_1$ in the case of quenched \cite{Kac02}, 2-flavor
\cite{Kac05} and 3-flavor \cite{Pet04} QCD. 

In the following we employ the parametrization obtained there for the
case $N_f=0$ --- since most lattice calculations of Euclidean correlators are
in quenched approximation --- but we shall also consider $N_f=2$, since in this
case lattice data for the free energy are available to us also at temperatures 
$0.76 \leq T/T_c\leq 1$ \cite{Kac05} and one can use them to study the behavior
of the correlation functions through the phase transition.
We refer the reader to Ref.~\cite{Alb05} for a detailed description of the
fitting procedure. 

Although in past work the free energy $F_1$ has been directly used as an input
for the $Q\bar{Q}$ potential energy, more recently it has been recognized that
the $Q\bar{Q}$ internal energy $U_1$ provides a more appropriate candidate.
Since the two quantities are connected by the well-known relation 
\begin{equation}
  F=U-TS,
\end{equation}
once a suitable parametrization of the temperature dependence for the free
energy has been obtained one can subtract the entropy contribution and get the
color-singlet internal energy as
\begin{equation}
\label{eq:U1}
  U_1 = -T^2\frac{\partial(F_1/T)}{\partial T}.
\end{equation}
The latter has been employed in Ref.~\cite{Alb05} as the effective $Q\bar{Q}$
potential in the Schroedinger equation, getting good agreement with the
charmonium dissociation temperatures extracted from lattice calculations of the
spectral functions. 

However, the use of the internal energy (\ref{eq:U1}) is not fully satisfactory
yet. From the theoretical point of view, as pointed out in Ref.~\cite{Won05}
and thoroughly discussed in Ref.~\cite{Won06a}, the internal energy of
Eq.~(\ref{eq:U1}) contains the genuine $Q\bar{Q}$ potential energy, but also a
contribution due to the variation of the gluon (and light quarks, for
$N_f\ne0$) internal energy in the presence of the $Q\bar{Q}$ pair.

Also the comparison with independent lattice results presents some
shortcomings. In fact, although it is quite successful in providing
dissociation temperatures in agreement with direct lattice estimates, its
asymptotic ($r\to\infty$) value --- employed to define the threshold energy
entering into the continuum part of the spectral function --- turns out to have
a sharp temperature dependence around the critical temperature. This in turn
is reflected, as we shall see in the next Section, into a temperature
dependence of the correlation functions which is not observed on the lattice.

\begin{figure}[t]
\begin{center}
\includegraphics[clip,width=0.5\textwidth]{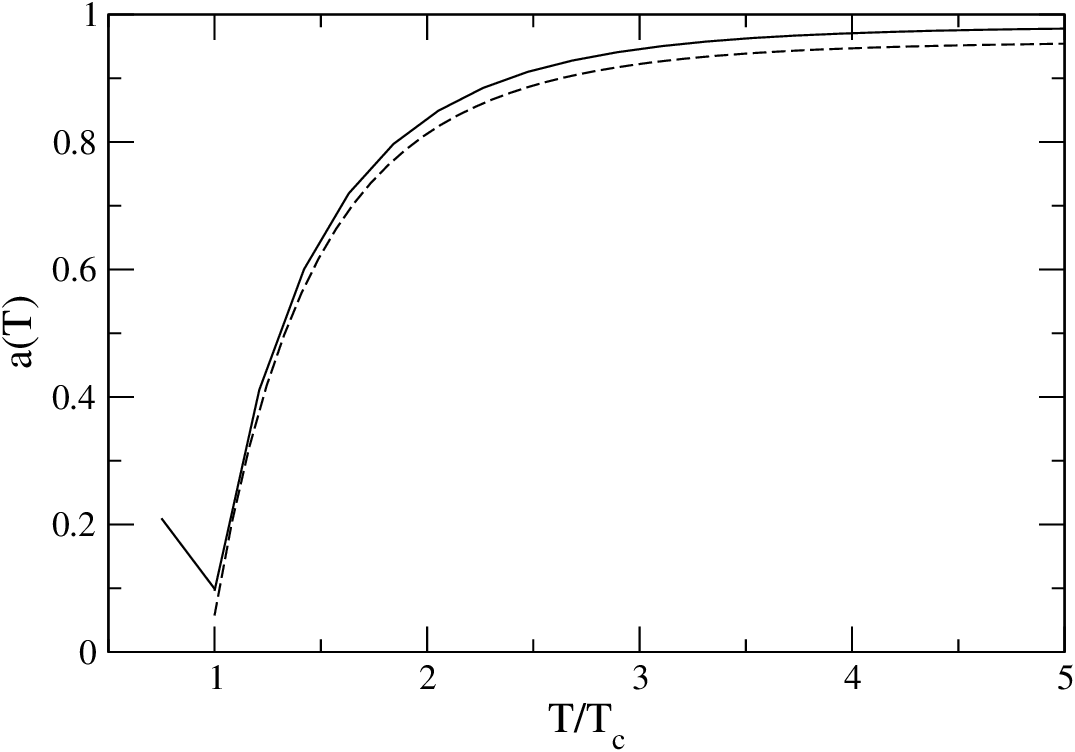}
\caption{ The ratio $a(T)=3p/\epsilon$ as a function of $T$ for $N_f=0$
  \protect\cite{Boy96} ($T\ge T_c$, dashed line) and $N_f=2$
  \protect\cite{Kar00} ($T\ge 0.76T_c$, solid line). }
\label{fig:aT}
\end{center}
\end{figure}

In this work we follow the procedure described in Ref.~\cite{Won05} in order to
separate the true $Q\bar{Q}$ internal energy, $U_1^{Q\bar{Q}}$, from the gluon
and light quark contribution. 
The effective potential entering into the Schroedinger equation is defined as
($F_1$ and $U_1$ are the same quantities considered in Eq.~(\ref{eq:U1})):
\begin{eqnarray}
\label{eq:V1}
  V_1(r,T) &\equiv& U_1^{Q\bar{Q}}(r,T)-U_1^{Q\bar{Q}}(r\to\infty,T) 
    \nonumber \\ 
  &=& f_F(T)\left[F_1(r,T)-F_1(r\to\infty,T)\right] +
    f_U(T)\left[U_1(r,T)-U_1(r\to\infty,T)\right],
\end{eqnarray}
where 
\begin{subequations}
\begin{eqnarray}
  f_F(T) &=& \frac{3}{3+a(T)}, \\
  f_U(T) &=& \frac{a(T)}{3+a(T)}
\end{eqnarray}
\end{subequations}
and
\begin{equation}
\label{eq:aT}
  a(T)=\frac{p}{\epsilon/3},
\end{equation}
$p$ and $\epsilon$ being the pressure and energy density of a homogeneous
system of quarks and gluons, respectively.
These thermodynamical quantities have been obtained in quenched \cite{Boy96} 
and 2-flavor \cite{Kar00} QCD as a function of temperature and we display in
Fig.~\ref{fig:aT} the resulting ratio (\ref{eq:aT}) in the range of
temperatures we are interested in. Note that the weight functions $f_F$ and
$f_U$ turn out to vary in the range $3/4\alt f_F\alt 1$ and $0\alt f_U\alt
1/4$, respectively, a fact which explains why the $Q\bar{Q}$ potential is
closer (but not identical) to the free energy $F_1$ than to the internal energy
$U_1$. 

\begin{figure}[t]
\begin{center}
\includegraphics[clip,width=0.6\textwidth]{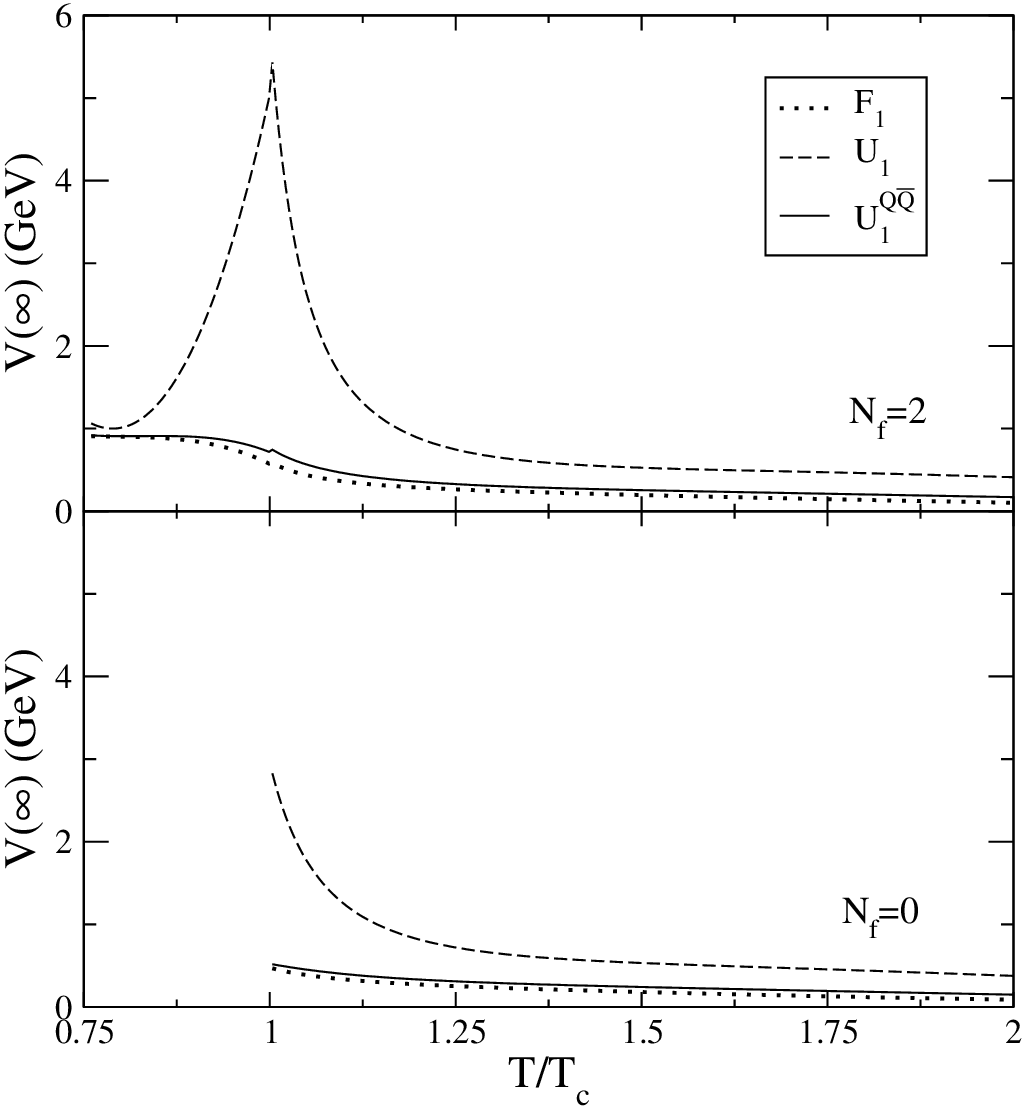}
\caption{ The asymptotic values of the lattice free energy
  \protect\cite{Kac05} (dotted), the total internal energy of
  Eq.~(\protect\ref{eq:U1}) (dashed) and the $Q\bar{Q}$ internal energy
  $U_1^{Q\bar{Q}}$ (solid) in 2-flavor and quenched QCD. }
\label{fig:U1asy}
\end{center}
\end{figure}

The asymptotic value of the $Q\bar{Q}$ internal energy,
$U_1^{Q\bar{Q}}(r\to\infty,T)$, is used to define the continuum threshold, 
$s_0(T)=2m+U_1^{Q\bar{Q}}(r\to\infty,T)$ and we compare it in
Fig.~\ref{fig:U1asy} to the asymptotic value of the internal energy as obtained
from Eq.~(\ref{eq:U1}), i.~e. without subtracting the gluon and light quark
contribution. 
The sharp variation of $U_1$ around $T_c$ --- which, as discussed in the next
Section, leads to correlation functions in contrast with lattice calculations
--- is evident, whereas $U_1^{Q\bar{Q}}$ displays a much smoother transition.

The effective potential based upon $U_1^{Q\bar{Q}}$ is less attractive than the
one based upon $U_1$ (but still more attractive than the one based upon $F_1$),
a fact which gives rise to slightly lower dissociation temperatures. 
Values for the latter in 
accord with the results from lattice spectral function studies can be obtained
by using quark masses slightly higher than the values of the Particle Data
Group listing \cite{PDG04}. For instance, in Refs.~\cite{Won06a} dissociation
temperatures compatible with the lattice phenomenology have been found using a
charm quark mass of $1.4\div 1.5$~GeV. 
Note that the lattice free energies employed to parametrize the effective
$Q\bar{Q}$ potential have been obtained for infinitely heavy quark mass and
that the lattice calculations of correlation functions use a variety of quark
masses, usually chosen in order to reproduce the mass of the charmonium
(bottomonium) ground state.

Here we solve the Schroedinger equation using an effective quark mass defined
as $\tilde{m}(T)=m+U_1^{Q\bar{Q}}(T)/2$, which, beside improving the binding of
the $Q\bar{Q}$ pair, can be naturally interpreted as a thermal mass.
Using this definition the continuum threshold is given as
$s_0(T)=2\tilde{m}(T)$. 
Our choice implies that the same thermal mass is employed both in the kinetic 
term of the effective hamiltonian and for the energy of the heavy quark at rest
in the plasma. In principle, nothing prevents these two masses from assuming 
different values. While the threshold mass can be extracted from the available
lattice data, in order to evaluate the thermal correction to the kinetic 
operator one should know the momentum dependence of the heavy quark self-energy
in the QGP. Lacking the above information, here we choose to ignore such a 
distinction. 
Note that in a non relativistic QED plasma a moving test charge receives a
positive thermal correction to its mass, which is slowly decreasing with the 
temperature~\cite{Ich73}. Our choice is in qualitative agreement with this
rigorous result obtained in another contest. 
On the other hand, it has been argued \cite{Moc06b} that the thermal mass
should not be used in the Schroedinger equation, since in the bound state the
heavy quarks should not feel the effect of the medium.
Yet, the quantitative differences between the two approaches are quite 
moderate, well below other uncertainties present in the problem.

\section{Results}
\label{sec:results}

\begin{table}[t]
\caption{\label{tab:dissT} Spontaneous dissociation temperatures (in units of
  $T_c$) of the lowest $c\bar{c}$ and $b\bar{b}$ states. The values in square
  brackets have been taken from Ref.~\protect\cite{Alb05} (see text); the ones 
  labeled with $(^*)$ have been obtained with a potential extrapolated beyond
  the temperature range of the lattice data ($T\alt2T_c$ for $N_f=2$). }
\begin{ruledtabular}
\begin{tabular}{|c|cc|cc|}
  & \multicolumn{2}{c|}{$N_f=0$} & \multicolumn{2}{c|}{$N_f=2$} \\
\hline
  & $m_c=1.4$ GeV & $m_c=1.6$ GeV & $m_c=1.4$ GeV & $m_c=1.6$ GeV \\
\hline
  $J/\psi,\eta_c$ & 1.40 & 1.52 & 1.45 & 1.59 \\
  $\chi_c$        & $<1$ & $<1$ & 1.00 & 1.00 \\
  $\psi'$         & $<1$ & $<1$ & 0.98 & 0.99 \\
\hline
\hline
  & $m_b=4.3$ GeV & $m_b=4.7$ GeV & $m_b=4.3$ GeV & $m_b=4.7$ GeV \\
\hline
  $\Upsilon,\eta_b$ & 2.96 [4.5] & 3.18 & 3.9(*) [6.7(*)] & 4.4(*) \\
  $\chi_b$          & 1.13 [1.55]& 1.15 & 1.15 [1.63]  & 1.17 \\
  $\Upsilon'$       & 1.12 [1.40]& 1.14 & 1.13 [1.43]  & 1.15 \\
\end{tabular}
\end{ruledtabular}
\end{table}

In this Section we use the effective potentials of Eq.~(\ref{eq:V1}) to
calculate spectral and correlation functions by solving the Schroedinger
equation
\begin{equation}
  \left[-\frac{\bm{\nabla}^2}{\tilde{m}} + V_1(r,T)
  \right]\psi_\varepsilon(\bm{r},T) = \varepsilon(T)
  \psi_\varepsilon(\bm{r},T), 
\end{equation}
for the $Q\bar{Q}$ eigenvalues $\varepsilon(T)$ and the eigenfunctions
$\psi_\varepsilon(\bm{r},T)$. 
The dissociation temperatures for the lowest $c\bar{c}$ and $b\bar{b}$ states
are shown in Tab.~\ref{tab:dissT}. 
Since the potential is spin-independent the two $S$-states, pseudoscalar and
vector (or $P$-states, scalar and axial-vector) are degenerate.

With respect to the previous findings with the full internal energy
\cite{Alb05} one now observes a reduction of the dissociation
temperatures: in the $c\bar{c}$ channel the ground state melts around
$\approx1.5\div1.6~T_c$, whereas the excited states already around the critical
temperature; in the $b\bar{b}$ channel the ground state melts above $3 T_c$ and
the excited states around $\approx1.1\div1.2~T_c$.
For comparison, in the Table we also report (in square brackets) the
dissociation temperatures obtained in Ref.~\cite{Alb05} using the full internal
energy, in the cases where the same quark mass had been employed (bottomonium
for $m_b=4.3$~GeV).

\begin{figure}[p]
\begin{center}
\includegraphics[clip,width=0.42\textwidth]{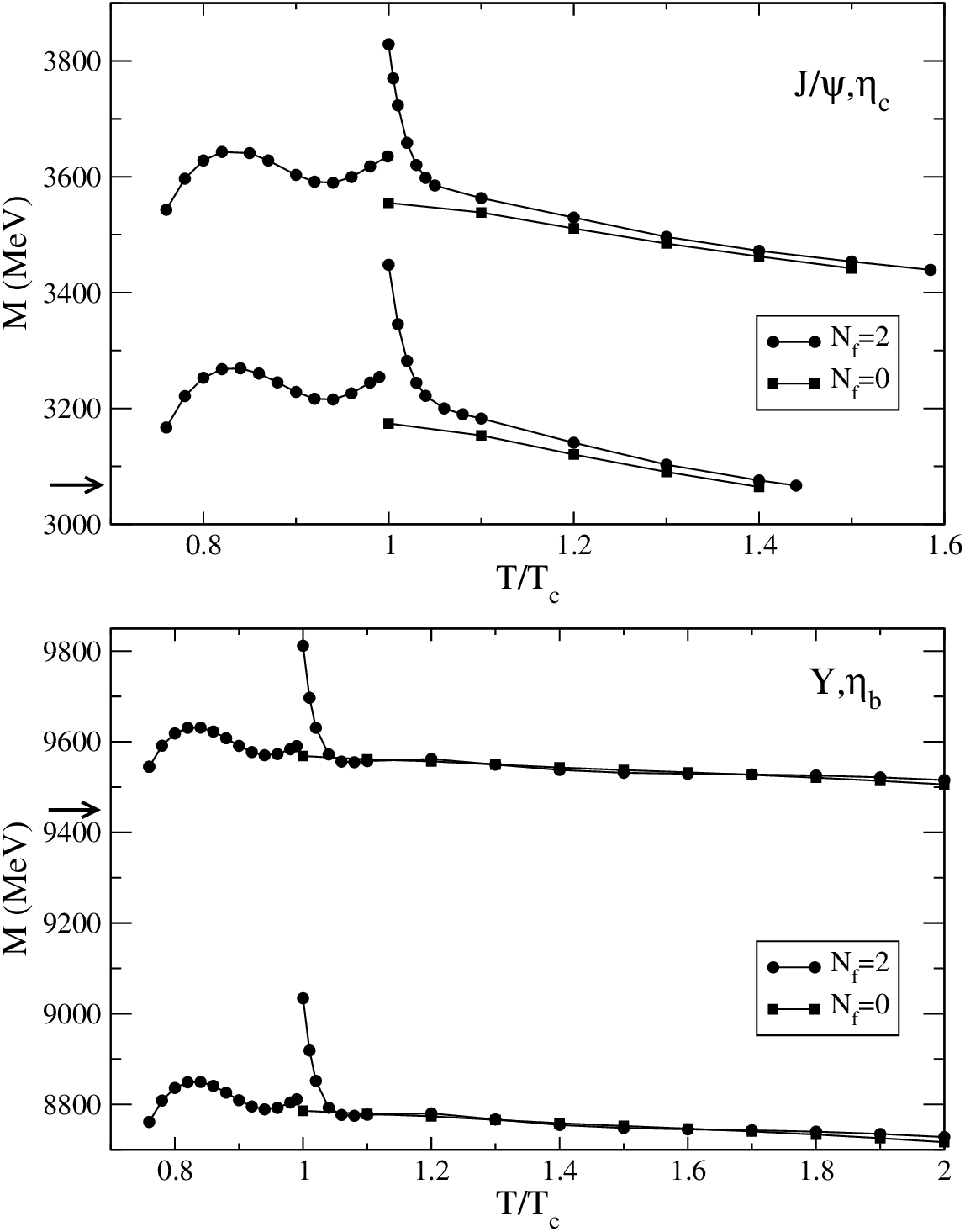}
\caption{ Mass as a function of temperature of the lowest $S$-wave $c\bar{c}$
  (upper panel) and $b\bar{b}$ (lower panel) states obtained from the solution
  of the Schroedinger equation. The upper curves are for a quark mass
  $m_c=1.6$~GeV ($m_b=4.7$~GeV), the lower ones for a quark mass $m_c=1.4$~GeV 
  ($m_b=4.3$~GeV). The arrows point to the value of the mass at $T=0$.}
\label{fig:Smasses}
\bigskip
\includegraphics[clip,width=0.42\textwidth]{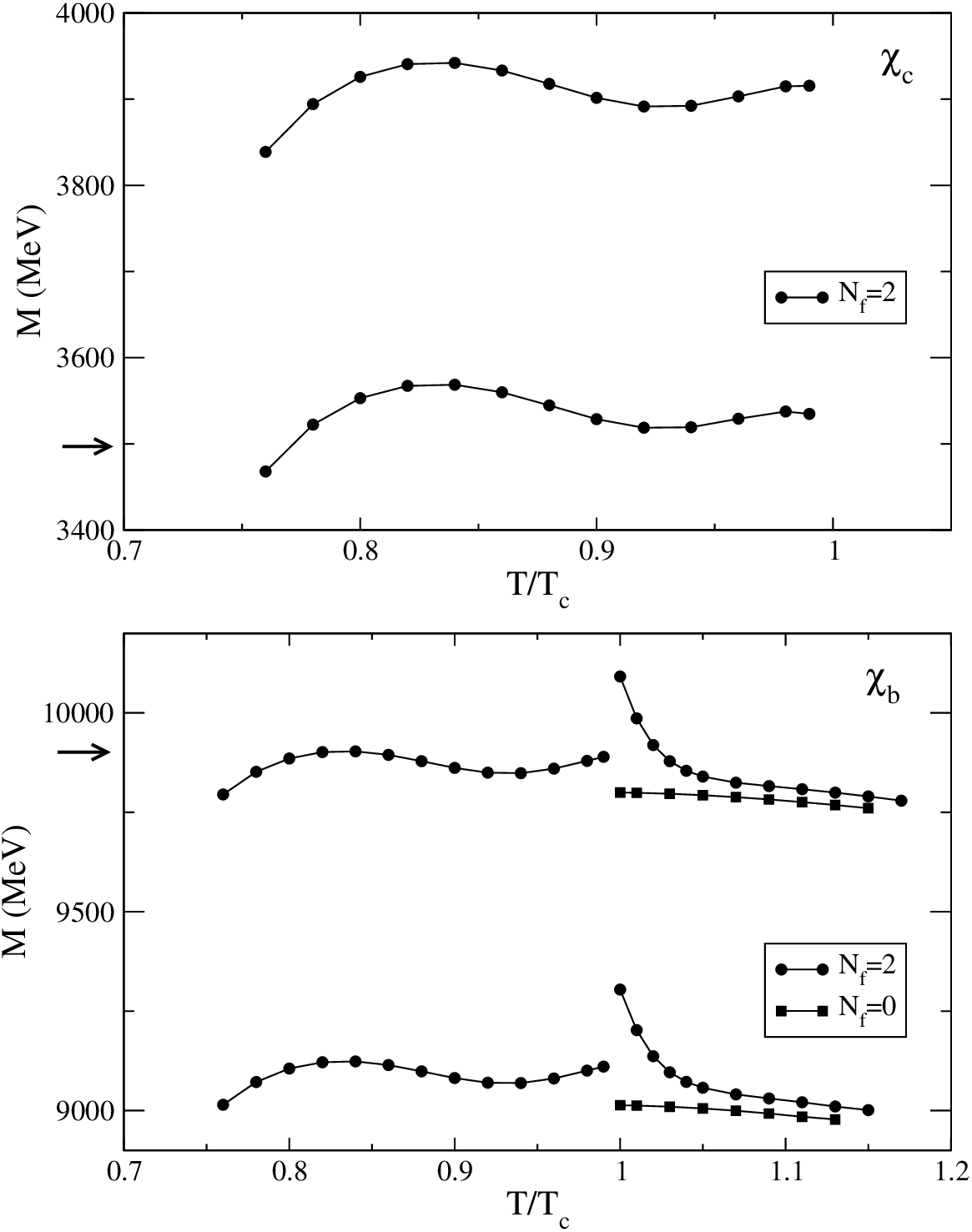}
\caption{ As in Fig.~\protect\ref{fig:Smasses}, but for the lowest $P$-wave
  states. }
\label{fig:Pmasses}
\end{center}
\end{figure}

In Figs.~\ref{fig:Smasses} and \ref{fig:Pmasses} we show the mass, 
$M=2m+U_1^{Q\bar{Q}}(r\to\infty,T)+\varepsilon(T)$, of the lowest
$S$-wave and $P$-wave states, respectively, as a function of temperature.
Apart from a narrow range of temperatures around $T_c$, the variation of the
masses with $T$ is quite moderate, the difference between the maximum and
minimum values being around $150\div200$~MeV of the $S$-states and
$\approx100$~MeV for the $P$-states. This, for instance, amounts to roughly
$5\div6$\% (2\%) of the $J/\psi$ ($\Upsilon$) mass.

A slight decrease of the $\eta_c$ mass in the range of temperatures explored
here has also been observed in some lattice spectral studies~\cite{Iid06}. 
Note that a more pronounced softening of the $J/\Psi$ peak (here degenerate
with $\eta_c$) as the 
temperature increases has been attributed by some authors to the presence in 
the vector channel of a transport contribution to the spectral function, not
resolved by the MEM procedure, but nevertheless tending to move strength 
towards lower energies~\cite{Jak06}.

Just above $T_c$ one observes a slight peak in the masses for the case
of 2-flavor QCD. Note that this is the range of temperatures where the
$T$-dependence of the parameters employed to fit the lattice free energies is
stronger \cite{Alb05}, so that it might just signal the inadequacy of the
parametrization at the critical temperature. It is however curious that the
same behavior does not appear for $N_f=0$, where the same parametrization has
been employed. 

Note that the smooth temperature dependence of the masses of the $Q\bar{Q}$
states is strictly related to the smooth $T$-dependence of the asymptotic
$Q\bar{Q}$ internal energy, $U_1^{Q\bar{Q}}(\infty,T)$, obtained through the
procedure described in Section~\ref{subsec:effpot}. By employing, in the
definition of the bound state mass, the total internal energy $U_1(\infty,T)$,
one would get much higher (by a few GeV) masses around the critical
temperature (see Fig.~\ref{fig:U1asy}). 

\begin{figure}[t]
\begin{center}
\includegraphics[clip,width=0.7\textwidth]{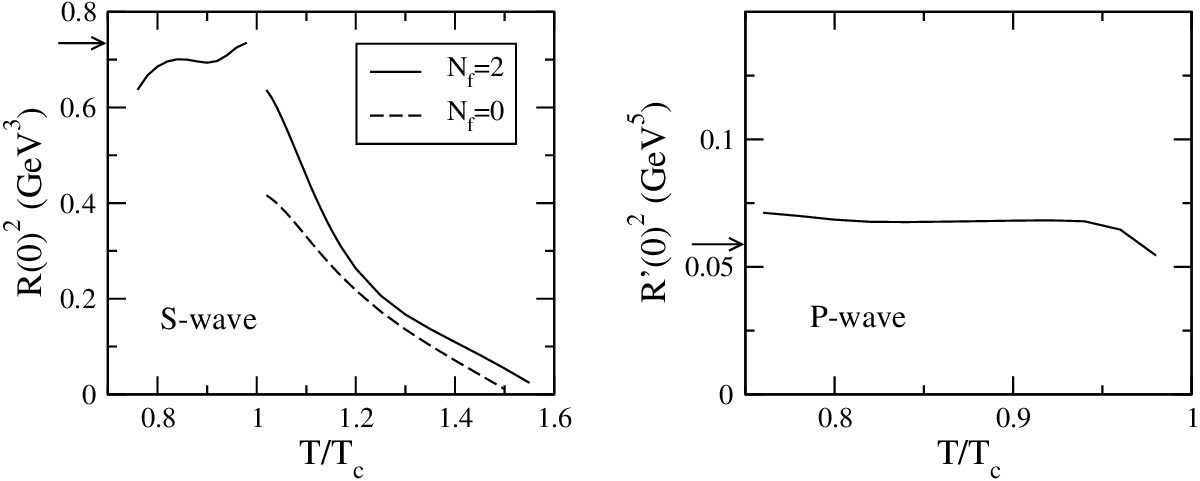}
\caption{ Squared value in the origin, for the $c\bar{c}$ system
  ($m_c=1.6$~GeV), of the $S$-wave radial wave function (left panel) and of the
  first derivative of the $P$-wave radial wave function (right panel), as a 
  function of the temperature, for quenched and unquenched QCD. The arrows
  point to the values at $T=0$, calculated with the Cornell potential of 
  Ref.~\protect\cite{Ful94}. }
\label{fig:R2SP}
\end{center}
\end{figure}

Other observables can be easily calculated in the potential model. As an
example, we show in Fig.~\ref{fig:R2SP}, for the $c\bar{c}$ system, the square
of the $S$-wave radial wave function, $R(0)^2$, and the square of the first
derivative of the $P$-wave radial wave function, $R'(0)^2$, calculated in the
origin. 
This quantities can be related to the quarkonium decay widths for different
processes (see, e.~g., Ref.~\cite{Bra05}), such as the leptonic decay rate of a
neutral vector meson \cite{Van67,Bar76} or $\chi$ decay into pairs of
pseudoscalar or vector mesons \cite{Dun80}.

As one can see from the figure, the values of $R(0)$ and, particularly, of
$R'(0)$ are quite stable below $T_c$ and very close to the values at $T=0$.
Above $T_c$ the $P$-wave states have melted, whereas $R(0)$ for the $S$-state
drops almost linearly up to the dissociation temperature.

\subsection{Spectral functions}

Here we would like to compare the spectral functions calculated in the
potential model to the ones extracted from the lattice euclidean correlators
using MEM. To make easier a direct comparison of the spectral functions it is
convenient to modify the potential model expression of Eq.~(\ref{eq:sigmaHPM})
in order to accommodate a width for the bound states.
Indeed, the width of the bound states at finite $T$ in the lattice spectral
functions contains a contribution due to the (possible) thermal broadening of
the states and a contribution due to the statistical uncertainties, the latter
apparently being dominant above $T_c$ because of the limited set of data points
\cite{Jak06}. 

A width can be easily included by substituting in Eq.~(\ref{eq:sigmaHPM}) the
delta function in the bound state sector with a Lorentz distribution:
\begin{equation}
\label{eq:sigmaHPMG}
  \sigma_H(\omega,T) = \sum_n F_{H,n}^2 \frac{1}{\pi} \frac{\Gamma_n/2}
    {(\omega-M_n)^2 + \Gamma_n^2/4} + \theta(\omega-s_0) F_{H,\omega-s_0}^2, 
\end{equation}
where $\Gamma_n$ represents the width of the state $n$.

\begin{figure}[p]
\begin{center}
\includegraphics[clip,width=\textwidth]{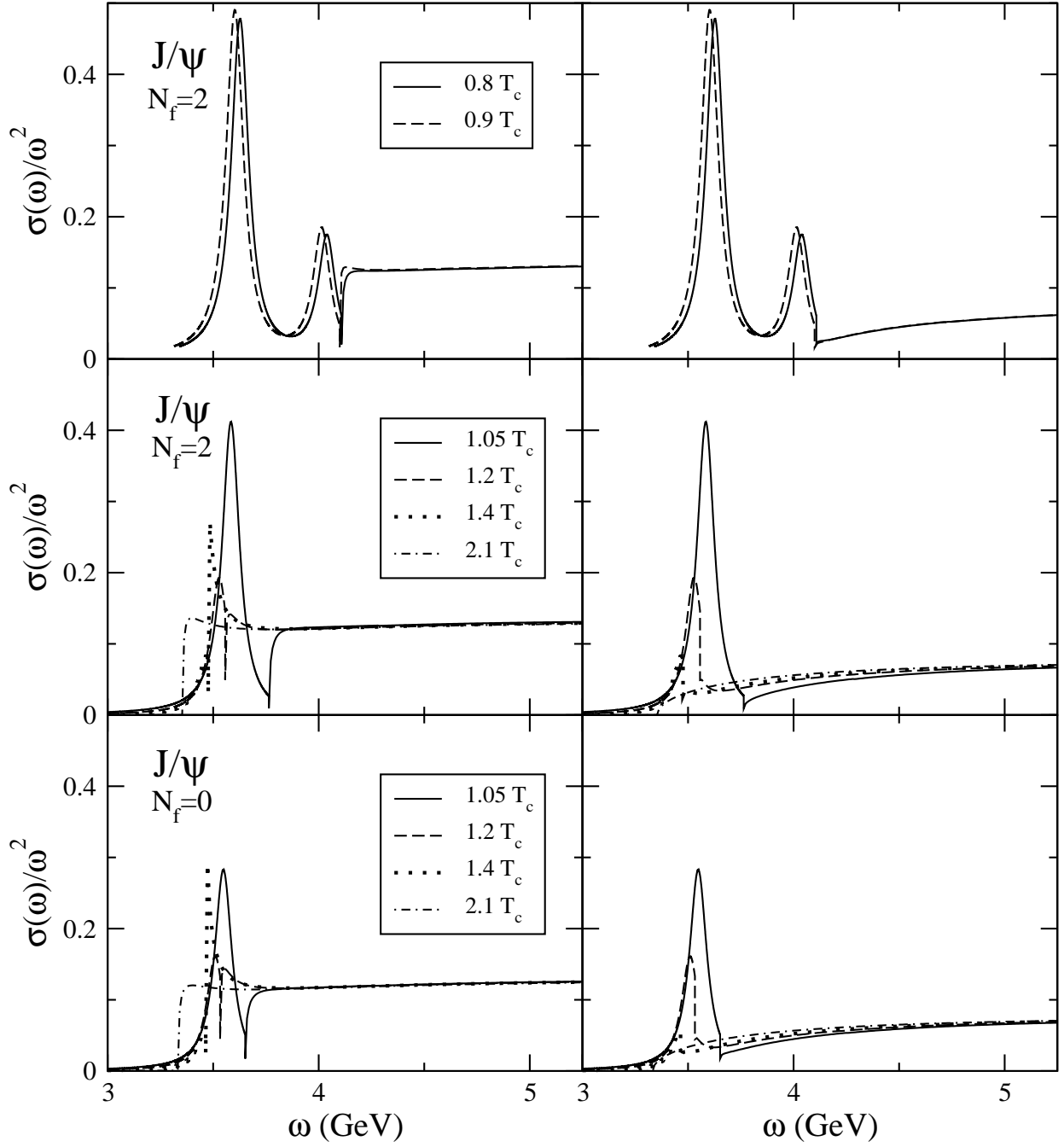}
\caption{ The $J/\psi$ spectral function divided by $\omega^2$ as a function of
  $\omega$ for $N_f=2$ and $N_f=0$ at several temperatures below and above
  $T_c$; in the left panels the continuum part of the spectrum comes from the
  solution of the Schroedinger equation, in the right panels from the
  perturbative expression. }
\label{fig:sigmaG100S}
\end{center}
\end{figure}

A word of caution is in order about the calculations at $T<T_c$ --- where we do
not deal with a quark-gluon plasma, but with a still confined system.
Although the use of a potential model for bound states is well assessed, at
least at $T=0$, the interpretation of the continuum spectrum is much more
questionable. Here we shall use both the spectral function calculated in
perturbative QCD, as usually done in the literature, and the one coming from
the potential model, in order to get a feeling about the model dependence of
the results. 
We would like however to stress that the aim of the model is to describe the
physics of quarkonia at $T>T_c$ and that calculations in the confined phase are
done in order to discuss the lattice results for the correlation functions, 
which are mainly available in the literature as ratios of functions above and 
below $T_c$ (see Sections~\ref{subsec:lcsp} and \ref{subsec:corrfunc}).
In potential model calculations, the reconstructed correlators have been
usually considered at zero temperature \cite{Moc06b,Won06b,Cab06}, adding to
the bound state contribution --- calculated with a phenomenological confining 
potential --- the perturbative QCD continuum. This choice obviously poses
another problem of consistency (besides the one, already mentioned, of the
consistency between the bound and the continuum spectra), since above $T_c$ one
is using a lattice-derived potential and at $T=0$ a phenomenological (fitted to
the data) one. 
Incidentally, the phenomenological parameters of the $T=0$ potential adopted in
the above works (i.e., the coefficients of the Coulomb and linear terms) turn 
out to be quite different with respect to the ones employed in normalizing the
free energies at short distances~\cite{Kac05}.
The use, also below $T_c$, of lattice-derived potentials should give one a
feeling about these model dependencies. 

\begin{figure}[p]
\begin{center}
\includegraphics[clip,width=0.45\textwidth]{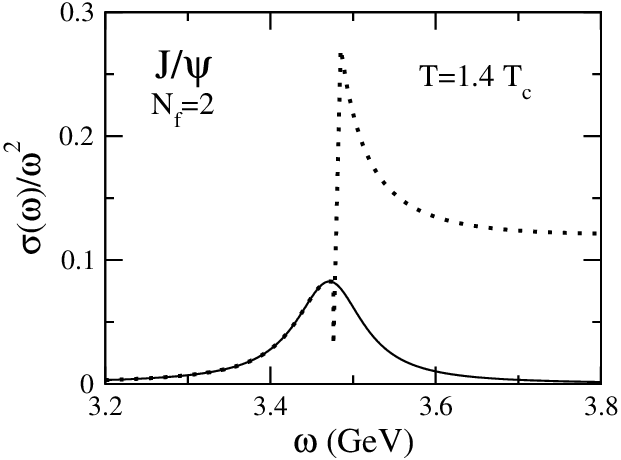}
\caption{ The $J/\psi$ spectral function divided by $\omega^2$ as a function of
  $\omega$ for $N_f=2$ and $T=1.4~T_c$ (dotted line). The bound state 
  contribution (solid line) is also shown separately. }
\label{fig:sigmaG100ST14}
\bigskip
\includegraphics[clip,width=0.7\textwidth]{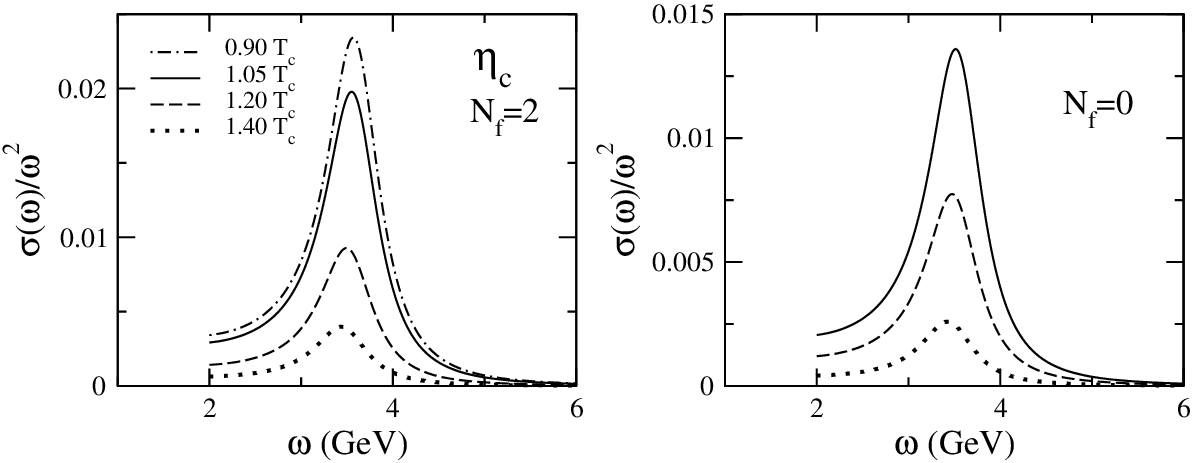}
\caption{ Bound state contribution to the pseudoscalar spectral function
  divided by $\omega^2$ for $N_f=2$ and $N_f=0$ at various temperatures. 
  A width $\Gamma_n=700$~MeV has been used. }
\label{fig:sigmaSbound}
\bigskip
\includegraphics[clip,width=0.7\textwidth]{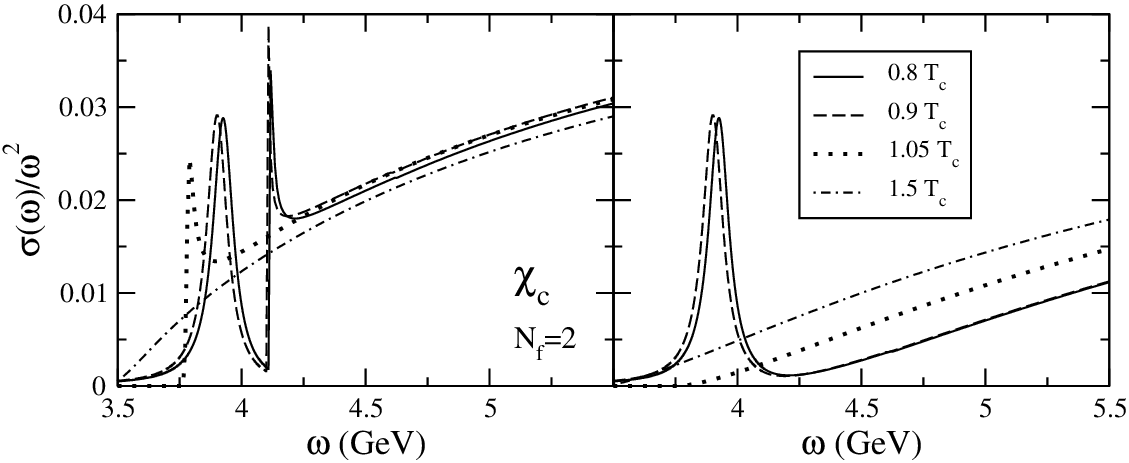}
\caption{ The $\chi_c$ spectral function divided by $\omega^2$ as a function of
  $\omega$ for $N_f=2$ at several temperatures below and above
  $T_c$; in the left panel the continuum part of the spectrum comes from the
  solution of the Schroedinger equation, in the right panel from the
  perturbative expression. }
\label{fig:sigmaG100P}
\end{center}
\end{figure}

In Fig.~\ref{fig:sigmaG100S} we display the $c\bar{c}$ spectral function for
the $S$-states for quenched and unquenched QCD, both below and above $T_c$
(since we neglect the hyperfine spin-spin interaction the $J/\psi$ and $\eta_c$
states are degenerate and the corresponding spectral functions differ for
trivial factors). In the left hand panels we employ --- for the continuum part
of the spectral function --- the solution of the Schroedinger equation
(Eq.~(\ref{eq:sigmaHPMG}), see the discussion in Section~\ref{subsec:lcsp}), 
whereas in the right hand panels we employ the perturbative expression (see 
Eq.~(\ref{eq:sigmaHPMA})). In all cases a width $\Gamma_n=100$~MeV has been 
used for the discrete states.
Here and in the following calculations we employ $m_c=1.6$~GeV and
$m_b=4.7$~GeV for the quark masss.

In this figure one can note a marked peak of constant strength and almost fixed
position below $T_c$, corresponding to the $J/\psi$ state, and a second smaller
peak, corresponding to its first radial excitation. Above $T_c$ only the 
ground state peak survives and its position moderately moves (consistently with
the mild mass shift displayed in Fig.~\ref{fig:Smasses}), whereas its strength
is gradually decreasing as one approaches the dissociation temperature.
There is no significant difference between $N_f=0$ and $N_f=2$. 

Above $T_c$ the bound state peak is clearly visible especially in the right
hand panels, where the perturbative continuum spectral function has been
employed. 
The continuum contribution calculated from the Schroedinger equation,
on the other hand, shows a resonant part when the bound state energy is
approaching zero and this resonant contribution tends to dominate over the
bound state peak (see Fig.~\ref{fig:sigmaG100ST14}, where, as an example, the
spectral function at $T=1.4~T_c$ is magnified around the threshold energy,
separating the bound state and continuum contributions). 

In order to show more clearly the evolution of the bound state above $T_c$ ---
and to make easier a comparison with the lattice results --- we display
in Fig.~\ref{fig:sigmaSbound} only the bound state part of the $S$-wave
spectral function using a larger effective width, comparable to the one of the
lattice charmonium spectral functions calculated in Refs.~\cite{Dat04,Jak06}
($N_f=0$) and \cite{Aar06} ($N_f=2$). As one can see the pattern as a function
of the temperature for the position of the bound state appears to be similar to
the lattice results. 

The temperature dependence of the strength of the peak, instead, seems to
differ from the lattice outcome. In Refs.~\cite{Dat04,Jak06}, the $S$-wave
spectral functions have also been extracted at various temperatures using the
same number of data points, in order to compare results affected by the same
statistical uncertainties: at least for the pseudoscalar state (the vector
channel being more uncertain) it has been found that the spectral function
maintains its strength up to 1.5~$T_c$. This result is, of course, affected by
large errors and also depends upon the reliability of the MEM procedure.
On the other hand, the potential model calculation yields a practically
constant strength up to $T_c$ and then a practically linear decrease, as it
could be inferred also from Fig.~\ref{fig:R2SP}.

In Fig.~\ref{fig:sigmaG100P} we show the spectral function for the $P$-states
in the case of unquenched QCD, using, for illustration, a width of 100~MeV
(again, we neglect hyperfine splitting of scalar and axial-vector states).
The continuum contribution in the left panel comes from solving the
Schroedinger equation, in the right panel from the perturbative calculation.
Also in this channel one can note, below $T_c$, a marked peak of constant
strength and approximately fixed position, corresponding to the $\chi_c$ state.
No bound states are present above $T_c$: the peak appearing in the left panel
at $T=1.05T_c$ is actually a resonance in the continuum contribution. Also the
narrow peaks visible at higher energy for $T<T_c$ are embedded in the
continuum. 
The situation of the $P$-wave lattice spectral functions is less clear, since
this channel is much harder to study on the lattice. All the studies
\cite{Dat04,Jak06,Aar06} of the $c\bar{c}$ system observe a strong
modification of the spectral functions above $T_c$ and this is usually
interpreted as a signature of dissociation of the $P$-wave states. However,
above $T_c$ some strength is observed at very low energies, especially in the
scalar channel, well below the expected continuum threshold and there is no
physical explanation for these findings.

\begin{figure}[t]
\begin{center}
\includegraphics[clip,width=0.7\textwidth]{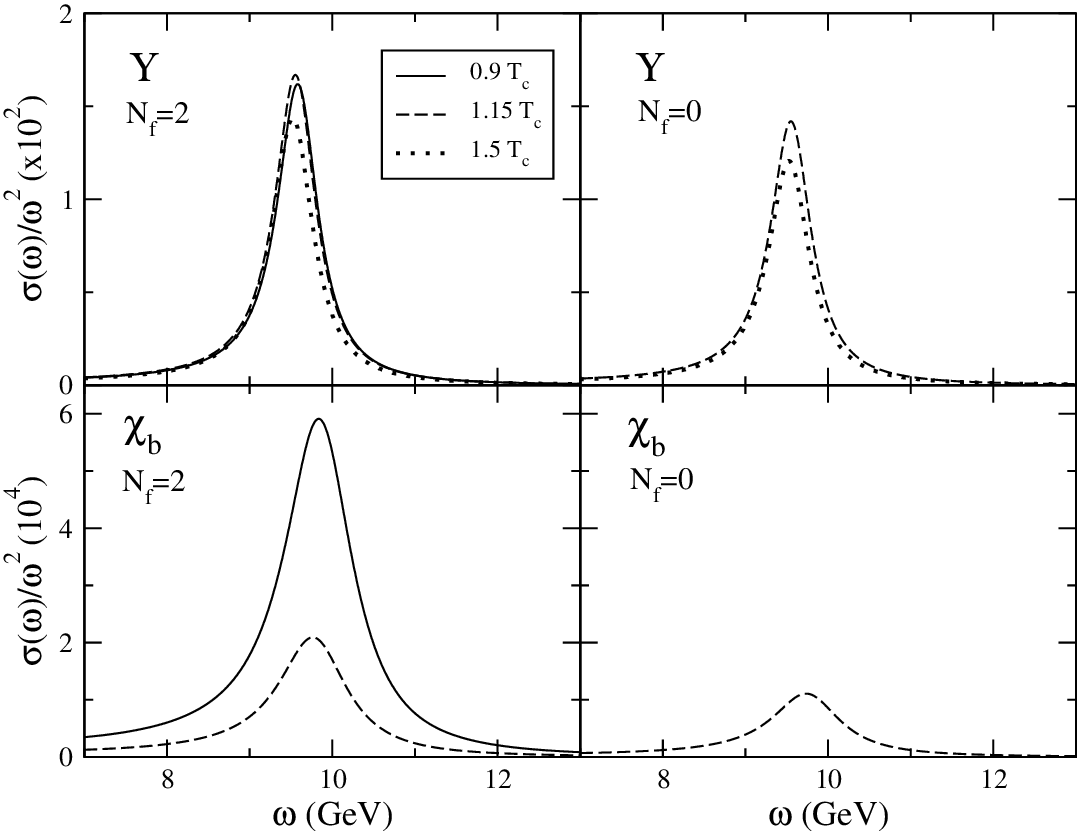}
\caption{ Bound state contribution to the spectral function divided by
  $\omega^2$ for $b\bar{b}$ states at temperatures below and above $T_c$. 
  $N_f=2$ in the left panels and $N_f=0$ in the right panels.
  The widths are 650~MeV ($\Upsilon$) and 1~GeV ($\chi_b$)
  \protect\cite{Pet05,Dat06}. }
\label{fig:sigmaSPbound}
\end{center}
\end{figure}

Finally, in Fig.~\ref{fig:sigmaSPbound} we display the bound state part of the
$S$- and $P$-wave spectral functions for $b\bar{b}$ states, in the range of
temperatures studied in Refs.~\cite{Pet05} and \cite{Dat06}.
The $\Upsilon$ state appears to be stable both in position and strength over
the whole range of temperatures, whereas the $\chi_b$ state gets strongly
modified approaching the dissociation temperature (see Table~\ref{tab:dissT}).
The stability in the position of the $\Upsilon$ and the dissociation of the
$\chi_b$ at a temperature $T\cong1.1\div1.2 T_c$ is in accord with the results
from lattice calculations; the latter, however, are still preliminary and one
cannot draw from them any conclusion about the evolution of the strength with
$T$. We do not show the continuum part of the spectrum, since it is very
similar to the charmonium case.

To summarize our findings for the $Q\bar{Q}$ spectral functions, we can state
that the bound state part of the potential model spectrum appears to be
consistent with the lattice results as far as the evolution with $T$ of the
position of the $c\bar{c}$ and $b\bar{b}$ states is concerned.
On the other hand, both the evolution of the bound state strength and the shape
of the continuum contribution --- either perturbative or calculated in the
potential model --- show major differences from the lattice outcome.

\subsection{Correlation functions}
\label{subsec:corrfunc}

As we have already mentioned above, the reliability of the MEM procedure is
still under discussion. In order to reveal the temperature dependence of the
spectral function, it has been proposed to consider the ratio between the 
euclidean correlator at a given temperature and the reconstructed correlator,
which contains the spectral function at a reference temperature (see
Section~\ref{subsec:lcsp}).

In lattice calculations, this has been done using as reference temperature both
some finite $T<T_c$ \cite{Dat04} and $T=0$ \cite{Jak06}. In potential model
calculations, the spectral function at $T=0$ --- modeled as in
Eq.~(\ref{eq:sigmaHPMA}) by using a phenomenological potential for the bound
states and the perturbative QCD contribution for the continuum states --- has 
been employed so far as a reference \cite{Moc06b,Cab06}. However, such a 
spectral function is not only affected by the inconsistency between the bound 
and the continuum states, but it also strongly depends upon the threshold
energy $s_0$, which has to be treated as a parameter.
Moreover, employing, for the continuum contribution in the confined phase, the 
perturbative QCD spectral function is at least problematic.
Here we try to remove the first two sources of uncertainties by employing the 
spectral functions calculated from the lattice-generated potential at $T<T_c$.
Of course, the problem of interpretation of the continuum spectrum is still
there, but, as we shall see below, it is probably inessential.

\begin{figure}[t]
\begin{center}
\includegraphics[clip,width=0.7\textwidth]{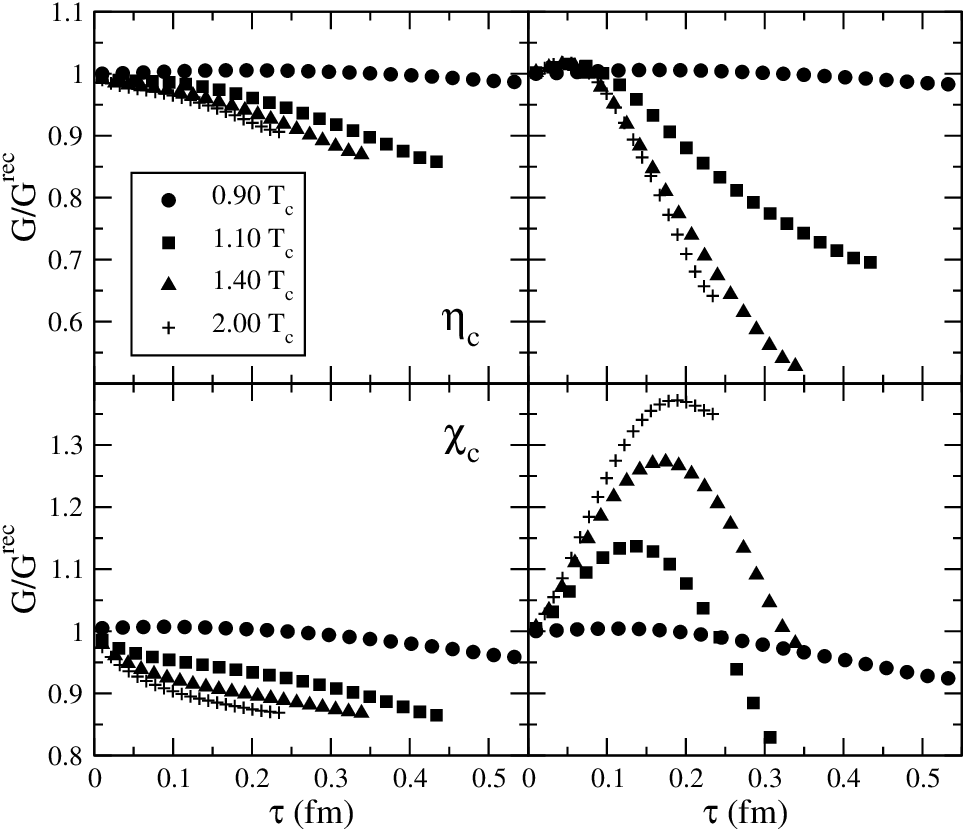}
\caption{ Ratio $G_H/G_H^\text{rec}$ for the pseudoscalar (upper panels) and 
  scalar (lower panels) charmonium states ($N_f=2$) at different
  temperatures, using the reconstructed correlator at $T=0.75 T_c$, as a
  function of the euclidean time $\tau$. In the left panels the continuum part
  of the spectrum comes from the solution of the Schroedinger equation, in the
  right panels from the perturbative expression. }
\label{fig:Gratio075}
\end{center}
\end{figure}

In Fig.~\ref{fig:Gratio075} we display the ratio $G_H/G_H^\text{rec}$ for the
pseudoscalar and scalar charmonium states at different temperatures, using the
reconstructed correlator at $T=0.75 T_c$. Both the continuum spectra generated
by solving the Schroedinger equation (left panels) and by using the
perturbative expressions (right panels) are employed.

Let us first analyze the ratio at $T<T_c$. Here it turns out to be always very 
close to one, not only for the case shown in the figure, but over the whole
range of temperatures available to us ($0.75<T/T_c<1$). It strongly resembles
the ratio of the correlators measured on the lattice \cite{Dat04}: even the
small decrease at large $\tau$'s in the scalar channel seems to be correctly
reproduced. Note that this outcome simply reflects the temperature independence
of the spectral functions below $T_c$, as it was apparent in
Fig.~\ref{fig:sigmaG100S}. This evolution with the temperature in our model
stems from the mild dependence upon $T$ of both the effective potential and the
threshold energy, which, in turn, is due to the procedure of subtraction of the
gluon and light quark contributions from the $Q\bar{Q}$ internal energy (see 
Section~\ref{subsec:effpot}). The same quantities calculated using directly the
internal $Q\bar{Q}$ energy as the effective potential would yield a strong 
$T$ dependence already below $T_c$ (see, e.~g., Fig.~\ref{fig:U1asy}).

Above $T_c$, the ratio in the pseudoscalar channel looks qualitatively similar
to the one measured on the lattice \cite{Dat04,Jak06} --- at variance with the
result of Refs.~\cite{Moc06b,Cab06} --- but there are important differences.
In the lattice results, the ratio remains close to one up to the dissociation
temperature and then it gradually decreases; here, the departure from one
occurs as soon as the temperature grows beyond $T_c$ and reflects the rapid
change in the threshold energy visible in Fig.~\ref{fig:U1asy}. After that
the smooth change in the threshold energy and the reduction of the bound state
strength as $T$ increases give rise to the moderate decrease of
$G_H/G_H^\text{rec}$ observed in Fig.~\ref{fig:Gratio075}. 

In the scalar channel the situation is rather different: the ratio calculated
using the perturbative continuum grows above one and then drops down, whereas
in the case of the potential model continuum it is always lower then one.
Both results are at variance with the lattice ones, where the ratio is
uniformly growing with $\tau$. The reason for the different behavior employing
either the perturbative or the ``interacting'' continuum can be reconducted to
the presence, in the latter case, of a strong resonance just above the
threshold at $T<T_c$ (see Fig.~\ref{fig:sigmaG100P}).

As one can see, differences in the outcome from the different models employed
here and in other works \cite{Moc06b,Cab06} can in general be understood in
terms of differences in the treatment of the continuum contribution to the
spectral functions (its form and/or the threshold energy).
Hence, the ratio $G_H/G_H^\text{rec}$ appears to be a sensitive observable in
discriminating among different models.
However, so far no model calculation has been able to reproduce the lattice
results for every channel.

\begin{figure}[t]
\begin{center}
\includegraphics[clip,width=0.8\textwidth]{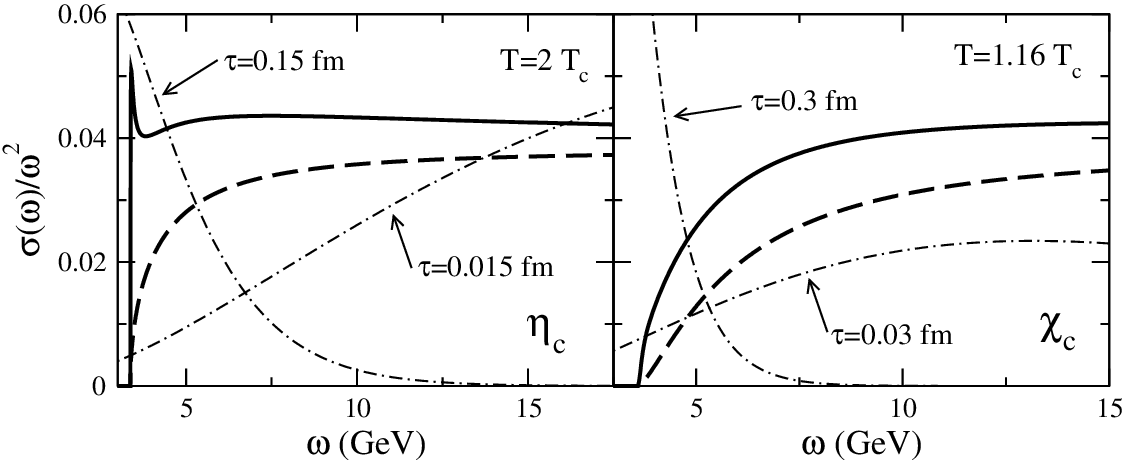}
\caption{ Spectral functions (divided by $\omega^2$) for $N_f=0$ in the
  pseudoscalar ($T=2 T_c$) and scalar ($T=1.16 T_c$) channels as a
  function of the energy; both the potential model (solid line) and the
  perturbative QCD (dashed line) results are displayed.
  The dot-dashed lines represent the factor $\omega^2 K(\tau,\omega,T)$ 
  (in arbitrary units) for two values of $\tau$. }
\label{fig:sigmaK}
\end{center}
\end{figure}

In order to understand the reason for this discrepancy, we display in
Fig.~\ref{fig:sigmaK} the spectral functions (divided by $\omega^2$) in the
pseudoscalar and scalar channels at a fixed temperature. In the same plots we
also show (dot-dashed lines) the factors $\omega^2 K(\tau,\omega,T)$ by which
the spectral function $\sigma_H(\omega,T)/\omega^2$ should be multiplied in
order to get the integrand yielding the euclidean correlator of
Eq.~(\ref{eq:GHs}). The factors $\omega^2 K(\tau,\omega,T)$ are shown for two
values of $\tau$, namely a small one --- where the ratio $G_H/G_H^\text{rec}$
goes to one both in lattice measurements and in model calculations --- and a
large one --- where the effects of temperature on the ratio are stronger.

As one can see from the figure, at small $\tau$'s the weighting factor
$\omega^2 K$ is a rather smooth function covering a wide range of energies and 
the corresponding correlator gives, roughly speaking, a measure of the total
strength associated to the spectral function. On the other hand, for large
values of $\tau$ the weighting factor $\omega^2 K$ is a rapidly dropping
function of the energy, selecting only the low energy part of the spectral
function (this is, of course, a nice feature for discriminating among different
models). 

The discrepancy with the lattice results can be understood by comparing, e.~g.,
Fig.~\ref{fig:sigmaK} with Fig.~17 (pseudoscalar case) and Fig.~19 (scalar
case) of Ref.~\cite{Jak06}. 
In the pseudoscalar case, at $T=2 T_c$ no bound state is present in our
potential model calculation. The lattice estimates at this temperature show a
strong dependence upon the default model used to extract the spectral function,
yielding either a significant change with respect to $T=0$ (see Fig.~17 of
Ref.~\cite{Jak06}) or almost no temperature dependence. In any case, the small 
reduction ($\approx5\%$) observed in the ratio of correlators is due to
(moderate) differences between the $T=0$ and $T=2 T_c$ spectral functions at
energies $\lesssim10$~GeV, which are dominated by the lowest energy peak
visible in Fig.~17 of Ref.~\cite{Jak06}. Our potential model calculation (solid
line in Fig.~\ref{fig:sigmaK}) also presents a peak (a resonance in the 
continuum spectrum) --- at variance with the perturbative QCD spectrum (dashed
line) --- and this explains the similar behavior of $G_H/G_H^\text{rec}$ in our
calculation and on the lattice. 

In the scalar case, on the other hand, there are no resonances in the
potential model calculation and no strength is present at low energy 
($\lesssim 4$~GeV) --- whereas some strength appears at low energy in the
lattice result (Fig.19 of Ref.~\cite{Jak06}) --- and this explains the
different behavior of $G_H/G_H^\text{rec}$ for this channel in our calculation
and on the lattice.  

We would like to stress that this explanation for the discrepancy between
potential model and lattice correlators does not rely upon the MEM-based
lattice spectral functions, which have been used only to illustrate our
argument. For instance, the exact lattice spectral function in the scalar
channel at $T=1.16~T_c$ may not have the form of the MEM-based ones (Fig.~11 of
Ref.~\cite{Jak06}), but it must have, in the low energy region, more strength
than the $T=0$ spectral function in order to yield a correlator twice as large
as the reconstructed one.

This analysis --- together with the direct comparison of the lattice and
potential model spectral functions done in the previous Section --- shows that 
the discrepancies between the two approaches are mostly located in the
continuum part of the $Q\bar{Q}$ spectrum.
Indeed, in this energy domain, the lattice spectral functions (which provide 
useful information on the properties of the ground state) suffer strong 
limitations, due to the inability of resolving excited/resonant states, the 
presence of unphysical peaks related to lattice artifacts and the bad
asymptotic high energy behavior. Note, in particular, that the presence of
unphysical peaks is not related to the MEM procedure, since they appear also in
the infinite temperature limit \cite{Kar03}.
Hence the ratio $G_H/G_H^\text{rec}$ may not be a good candidate to check the
consistency of the two approaches (potential models vs. lattice studies).

\section{Summary and conclusions}
\label{sec:concl}

We have given a comprehensive treatment of quarkonia at finite temperature,
within the framework of a potential model.
We have constructed an effective $Q\bar{Q}$ potential as a linear combination
of the finite temperature $Q\bar{Q}$ internal and free energies --- in order to
separate the genuine $Q\bar{Q}$ energy from the gluon and light quark
contributions --- and we have used as input the $Q\bar{Q}$ free energies
obtained on the lattice from Polyakov loop correlators.

The effective potential that we have obtained yields dissociation temperatures
for the $c\bar{c}$ and $b\bar{b}$ systems that are in agreement with estimates
based on independent lattice studies of euclidean correlators and spectral
functions. Specifically, we have found that in the charmonium system only the
ground $S$-wave states ($J/\psi$ and $\eta_c$) survive up to
$\approx1.5\div1.6~T_c$, all the other states melting around the critical
temperature or below; in the bottomonium system, again only the ground $S$-wave
states ($\Upsilon$ and $\eta_b$) survive up to temperatures $\gtrsim3~T_c$,
while $\chi_b$ and $\Upsilon'$ melt around $\approx1.1\div1.2~T_c$.

In order to gain deeper insight into the comparison of results based upon the
potential model and lattice studies, we have also calculated the $c\bar{c}$ and
$b\bar{b}$ spectral functions over a wide range of temperatures.
To test the model dependence of the results, two models for the continuum
spectrum have been employed: one using the interacting solution of the
Schroedinger equation and one using the perturbative QCD spectrum.
The evolution with the temperature of the mass of the bound states has been
found in good agreement with the lattice results, while the $T$-dependence of
the strength of the bound states agrees only below $T_c$. Above $T_c$ the
lattice-based spectral functions show little temperature dependence of the
bound state strength, whereas in the potential model --- where the strength 
is proportional to the coupling $F_H^2$ --- the latter drops almost linearly. 
On the other hand, the continuum spectrum in the two approaches shows no 
resemblance at any temperature. 

We have also calculated in the potential model the $Q\bar{Q}$ euclidean
correlators, since these quantities are directly measured on the lattice and
are not affected by the uncertainties inherent to the procedure of extraction
of the lattice spectral functions. We have found good agreement with the
lattice results in the pseudoscalar channel and not in the scalar one:
we have shown that the agreement or disagreement between potential model and
lattice in the correlators is not driven by the bound state contributions, but
by the continuum spectrum. Since the latter is known to be strongly affected by
artifacts due to the finite size of the lattice, the ratio of euclidean
correlators, $G_H/G_H^\text{rec}$, does not appear to be appropriate for a test
of potential models vs. lattice calculations.

At the present stage, the only substantial conflict between potential model and
lattice 
predictions for quarkonium properties at finite temperature seems to be in the
$T$-dependence of the bound state strength from the critical temperature up to
the melting point, i.~e. a progressive drop in the potential model and ---
presumably, since lattice results are available only for a few temperatures ---
a rapid change at the dissociation point in the spectral studies on the
lattice. 
Lattice results for the $Q\bar{Q}$ spectral functions above $T_c$, however, are
still affected by uncertainties related to the MEM procedure and by large
statistical errors and new measurements with better statistics will probably
put this outcome on firmer grounds. 
On the other side, one has to gain deeper insight into the connection of the 
potential model with the underlying QCD at finite temperature, not only to
provide a solid and rigorous basis for the model, but also in order to extend
its applicability (e.~g, by including an imaginary part to describe the thermal
broadening of the quarkonium states).

\section{Acknowledgments}
A. B. would like to thank Della Riccia foundation for financial support.

\end{document}